\documentclass[aps,prb,twocolumn,floatfix,footinbib,showpacs,superscriptaddress]{revtex4-1}
\usepackage{graphicx}
\usepackage{amsfonts,amsmath,amssymb}
\usepackage{amsthm}
\usepackage{dsfont,bm}
\usepackage{color}
\usepackage{soul} 
\usepackage{amsbsy}
\usepackage[colorlinks=true,linkcolor=blue,pagecolor=blue,filecolor=blue,menucolor=blue,urlcolor=blue,citecolor=blue,anchorcolor=blue]{hyperref}%
\usepackage{times}
\usepackage{mathtools}
\usepackage{etex}
\usepackage{amsmath}
\usepackage{amstext}

\newcommand{\nunder}[2][5]{\mathrlap{\mkern\the\numexpr#1/2mu\relax\underline{\phantom{\mathrm{#2}\mkern-#1mu}}}#2}

\newcommand{\ua}{\uparrow}
\newcommand{\da}{\downarrow}

\begin{document}

\title{Control of superconducting pairing symmetries in monolayer black phosphorus}

\author{Mohammad Alidoust}
\affiliation{Department of Physics, K.N. Toosi University of Technology, Tehran 15875-4416, Iran}
\author{Morten Willatzen}
\affiliation{Beijing Institute of Nanoenergy and Nanosystems, Chinese Academy of Sciences,
No. 30 Xueyuan Road, Haidian District, Beijing 100083, China}
\affiliation{Department of Photonics Engineering, Technical University of Denmark, Kongens Lyngby, DK-2800, Denmark}
\author{Antti-Pekka Jauho}
\affiliation{Center for Nanostructured Graphene (CNG), DTU Nanotech, Technical University of Denmark, DK-2800 Kongens Lyngby, Denmark}

\date{\today}
\begin{abstract}
Motivated by recent experimental progress, we study the effect of mechanical deformations on the superconducting pairing symmetries in monolayer black phosphorus (MBP). Starting with phonon-mediated intervalley spin-singlet electron-electron pairing and making use of realistic band parameters obtained through first-principles calculations, we show that the application of weak mechanical strain in the plane of MBP can change the effective $s$-wave and $p$-wave symmetry of the superconducting correlations into effective $d$-wave and $f$-wave symmetries, respectively. This prediction of a change in the pairing symmetries might be experimentally confirmed through angular dependence high-resolution tunneling spectroscopy, the Meissner effect, and critical temperature experiments. The idea of manipulating the superconducting symmetry class by applying planar mechanical strain can be extended to other anisotropic materials as well, and may help in providing important information of the symmetries of the order parameter, perhaps even in some high-$T_c$ superconductors.
\end{abstract}
\pacs{74.78.Na, 74.20.-z, 74.25.Ha}
\maketitle

\section{Introduction}

Black phosphorus (BP) is a puckered orthorhombic semiconductor with a tuneable direct band gap, offering striking prospects and unprecedented opportunites to field-effect transistors, photoelectrics, and photonics devices\cite{kim2,Xia,Carvalho,Rudenko,wei,Li1,liu1,Wang2,Pereira}. The band gap can be tuned by mechanical strain, ranging from $\sim 1.5$~eV in a monolayer to $\sim 0.59$~eV in a five-stack layer\cite{J.Qiao,V.Tran,J.Kim,Koenig}. In Ref. \onlinecite{V.Tran}, it is argued that the optical band gap of monolayer BP is around 1.5 eV corresponding to a band gap of $\sim 2.3$~eV down-shifted by the exciton binding energy of $\sim 0.8$~eV. Furthermore, the mobility of BP at room temperature exceeds $ 10^3$~cm$^2$/Vs\cite{Li1,liu1} becoming comparable to that of a graphene, i.e., $\propto 10^4$~cm$^2$/Vs\cite{Han,Doganov,Zou1,Liu2}.

Raman spectroscopy is a powerful technique to study the phonon modes and the structural dynamics of materials\cite{Sugai}. Recently, this technique has been utilized to characterize BP in detail\cite{Yang,Wu,Ribeiro,Gupta,Chakraborty}. Polarized Raman spectroscopy with a complete angular dependence characterization allows for the fitting of the angular dependence of the spectra. The studies revealed that the Raman spectra of BP can be explained by the symmetry groups $\rm A_g$ ($s$ and $d_z$ orbitals) and $\rm B_{2g}$ ($d_{xy}$ orbital) with anomalous complex values of the Raman tensor elements\cite{Wu,Ribeiro,Yang}. The Raman active phonon modes originate from out-of-plane vibrations ($\rm A_g^1$) and atomic motions in the plane of BP ($\rm A_g^2$ and $\rm B_{2g}$)\cite{Wu,Ribeiro,Yang,Gupta,Chakraborty}.

Two-dimensional materials are a unique platform to host Dirac and Weyl fermions (pertaining to relativistic physics) and topologically protected phenomena. Therefore, the existence of superconductivity in two-dimensional materials is of fundamental interest due to the interplay of superconductivity, topology, and relativistic physics that potentially can result in groundbreaking phenomena, uncover novel particles such as Majorana fermions, and revolutionize future generations of computers\cite{Nayak2008RMP,Beenakker2013ARCM,ramon}. Black phosphorus under a pressure of $\sim 10$~$\rm GPa$ undergoes an orthorhombic to simple-cubic structural transition and experiences a superconducting phase transition with a critical temperature ($T_c$) of $\sim 4.8$~$\rm K$, whose origin is still unexplained but could be due to the electron-phonon coupling mechanism\cite{Kawamura1,Karuzawa2,Shirotani}. Several works have attempted to provide a theoretical description for the observed superconducting phase in BP\cite{Flores,Wang1,Livas,Q.Huang,Jun-JieZhang,Ge,Gao,YanqingFeng,R.Zhang,Yuan,Saberi,Szewczyk}. The effect of strain on the $T_c$ of BP was studied from first-principles using the Eliashberg spectral function\cite{YanqingFeng,Q.Huang,Jun-JieZhang,Ge,Gao}. An increase of $T_c$ from $\sim 3$~$\rm K$ to $\sim 16$~$\rm K$ was attributed to a $\rm B_{3g}$ phonon mode and it appeared that a biaxial strain is more efficient in enhancing $T_c$ than uniaxial strain\cite{Ge,Gao}. In a recent experimental attempt, a BP crystal was intercalated by several alkali metals (Li, K, Rb and Cs, Ca). It was found that superconductivity shows a universal critical temperature of $\rm 3.8\pm 0.1~K$, independent of the synthesized chemical composition\cite{R.Zhang}. This unusual finding has brought up controversies about the origin of the reported intrinsic superconductivity in BP obtained by doping\cite{Yuan}. The determination of dominant superconducting pairing type(s) in a material is a formidable task. This challenge arises even in the case of graphene. On the theory side, several different pairing types such as $s$-wave, $p+ip$, $d+id$, and $f$-wave pairings have been proposed but there is still no unique experimental determination of the predominant pairing symmetry\cite{Ma,Faye,Nandkishore,Kiesel,Uchoa}. Generally, on one hand, theory predictions depend on the strength of interactions, the pairing mechanism, and approximations made in a model. On the other hand, reaching perfect experimental conditions, as theories assume, has proven to be highly challenging. The critically adverse impacts include unwanted defects, structural imperfections, and detrimental impurities (when dopants are required) that are introduced during the synthesis of samples. Therefore, a detailed experimental determination of the actual structures, crystal symmetries, predominant interactions, and pairing mechanism(s) are critical information to identify the pairing(s) that is (are) the most energetically desirable in a given material. Nevertheless, theoretical studies of the nucleation of superconductivity in materials and possible pairing symmetries can facilitate the exploration of new superconductors with favorable properties, which is intriguing to not only the fundamental sciences but also functional applications in high-sensitive devices. For example, the prediction of topological superconducting phase and its related exotic particles using conventional semiconductors has driven an intense wave of experimental interests and advancements during past few years\cite{Nayak2008RMP,Beenakker2013ARCM,ramon}. A recent experiment on a Weyl semimetal, $\rm MoTe_2$, explored the implications of an external control on superconductivity by applying mechanical strain\cite{wylsuperc_exp1}. It was found that the critical temperature increases following a change in the crystal structure of $\rm MoTe_2$. This enhancement might be explained by a recent theory where the increase of $T_c$ is caused by the emergence of the type-II phase of Weyl semimetals\cite{MA_type2,shapiro,volovik}.

Motivated by recent experimental progress described above, here we use realistic values for the band parameters of monolayer black phosphorus (MBP) under the influence of strain and show that, remarkably, the exertion of a proper biaxial strain in the plane of MBP can convert the effective symmetry class of $s$-wave and $p$-wave superconducting correlations into $d$-wave and $f$-wave classes, respectively. To uncover the pairing symmetry change, starting with phonon-mediated intervalley spin-singlet electron-electron pairing we derive the anomalous Green's function $\hat{f}$ and examine the intervalley spin-singlet correlations as a function of $k_x$ and $k_y$. Our findings demonstrate that the pseudospin-triplet spin-singlet correlation is also nonzero with odd symmetry in parity. We extract the coefficients of momenta $k_x^2$ and $k_y^2$ as a function of applied strain and determine the desired range and type of strain for achieving the change of pairing effective symmetries. We also consider the exchange field interaction, ${\bf h}$, that can be provided experimentally by proximity coupling of MBP to a ferromagnet. In the presence of $\textbf{ h}$, the spin-triplet components of $\hat{f}$ can be nonzero. We find that the change of symmetry classes occurs for these now nonzero spin-triplet correlations too. We finally discuss how the predicted change of pairing symmetry classes can be detected by angular-dependence high-resolution tunneling-spectroscopy experiments, the Meissner effect, and may alter the superconducting critical temperature.

\section{Model and Results}

MBP in the presence of strain, $\varepsilon_{ii}$, is described by the following low-energy effective Hamiltonian\cite{Voon1,Voon2}: 
\begin{equation}\label{Hamil}
\begin{split}
H= \int &\frac{d\textbf{k}}{(2\pi)^2}\hat{\psi}^\dag_{\textbf{k}}H(\textbf{k})\hat{\psi}_{\textbf{k}}=\\=\int &\frac{d\textbf{k}}{(2\pi)^2}\hat{\psi}^\dag_{\textbf{k}}\sum_{i,j=x,y}\Big\{ \big [u_0+\alpha_i\varepsilon_{ii} +(\eta_j+\beta_{ij}\varepsilon_{ii})k_j^2\big]\tau_0 +\\+&\big[\delta_0+\mu_i\varepsilon_{ii} +(\gamma_j+\nu_{ij}\varepsilon_{ii})k_j^2\big ]\tau_x -\chi_y k_y\tau_y\Big\}\hat{\psi}_{\textbf{k}},
\end{split}
\end{equation}
where $x$ and $y$ label the coordinates of the plane of MBP. Here $\tau_{x,y}$ are the Pauli matrices in pseudo-spin space. In what follows, for biaxial strain both $\varepsilon_{xx}$ and $\varepsilon_{yy}$ are nonzero, while for uniaxial strain  only one of them is nonzero. To include an exchange field, $\textbf{ h}=(h_x,h_y,h_z)$, we invoke the real spin degree of freedom and assign three directions of spin to each valley by $h_x\sigma_{x}+ h_y\sigma_{y}+ h_z\sigma_{z}$ in which $\sigma_{x,y,z}$ are the Pauli matrices. Hence, in the presence of spin degree of freedom, the field operator associated with the above Hamiltonian reads $\hat{\psi}^\dag(\textbf{k})=(\psi_{A\ua}^\dag, \psi_{A\da}^\dag,\psi_{B\ua}^\dag, \psi_{B\da}^\dag)$; the pseudospins and spins are labeled by $AB$ and $\ua\da$, respectively. The band parameters of MBP, appearing in the Hamiltonian, are given in Table \ref{table}. These parameters are obtained through density-functional-theory computations and symmetry calculations\cite{Voon1,Voon2}.     

We begin with singlet superconductivity that can be described by 
\begin{equation}
\Delta^{AB}_{\ua\da}\langle\psi^\dag_{A\ua}\psi^\dag_{B\da}\rangle+\text{H.c.},
\end{equation}
that is, phonon-mediated intervalley electron-electron coupling with opposite spins. Note that, generally, other pairing amplitudes can be nonzero as well. Nevertheless, we assume that they can be ignored because the considered mechanism is most likely favored experimentally, as is also found for graphene\cite{beenakker}. Thus, in what follows, we simply set $\Delta^{AB}_{\ua\da}=\Delta$. In order to study the pairing symmetries that can develop in MBP, we define the following propagators\cite{abrikosov}:
\begin{subequations}
\begin{align}
 g_{\tau\sigma\tau'\sigma'}(t-t'; \mathbf{ r}, \mathbf{ r}') = - \Big\langle {\cal T}\psi_{\tau\sigma} (t,\mathbf{ r}) \psi_{\tau'\sigma'}^{\dag}(t',\mathbf{ r}')  \Big\rangle, \\\underline{g}_{\tau\sigma\tau'\sigma'}(t-t'; \mathbf{ r}, \mathbf{ r}') = - \Big\langle {\cal T} \psi^{\dag}_{\tau\sigma} (t,\mathbf{ r}) \psi_{\tau'\sigma'}(t',\mathbf{ r}')  \Big\rangle,
\\f_{\tau\sigma\tau'\sigma'}(t-t'; \mathbf{ r}, \mathbf{ r}') = + \Big\langle {\cal T} \psi_{\tau\sigma} (t,\mathbf{ r}) \psi_{\tau'\sigma'}(t',\mathbf{ r}')  \Big\rangle,
\\f^{\dag}_{\tau\sigma\tau'\sigma'}(t-t'; \mathbf{ r}, \mathbf{ r}') = + \Big\langle {\cal T} \psi^{\dag}_{\tau\sigma} (t,\mathbf{ r}) \psi^{\dag}_{\tau'\sigma'}(t',\mathbf{ r}')  \Big\rangle, 
\end{align}
\end{subequations}
where $t, t'$ designate the imaginary time, ${\cal T}$ is the imaginary-time-ordering operator, and $\psi_{\tau\sigma}$ are field operators. Here, indices $\tau\tau'$ and $\sigma\sigma'$ stand for the pseudospin and spin degrees of freedom, respectively. In the Nambu particle-hole space, the propagators are related to the Hamiltonian through:
\begin{equation}\label{Green1}
\begin{split}
\left( \begin{array}{cc}
-i\omega_n+\hat{H}(\mathbf{ r})& \hat{\Delta}(\textbf{r}) \\
\hat{\Delta}^\dag(\textbf{r}) & i\omega_n +\tau_y\sigma_y\hat{H}^*(\mathbf{ r})\tau_y\sigma_y
\end{array} \right)\check{g}(i\omega_n;\mathbf{ r},\mathbf{ r}')\\=\delta(\mathbf{ r}-\mathbf{ r}'),
\end{split}
\end{equation}
in which $\omega_n=\pi (2n+1)k_BT$ is the Matsubara frequency, $n\in \mathbb{Z}$, $k_B$ is the Boltzman constant, $T$ is temperature, and $\Delta({\mathbf r})$ determines the spatial dependence of superconducting gap. Also, $\hat{H}(\mathbf{ r})$ is obtained by replacing ${\bf k}\rightarrow -i\boldsymbol{ \nabla}$ in $H(\textbf{k})$. The matrix form of the propagators is represented by $\check{g}(i\omega_n;\mathbf{ r},\mathbf{ r}')$ which is expressed as:  
\begin{equation}
\check{g}(i\omega_n;\mathbf{ r},\mathbf{ r}')=\left(  \begin{array}{cc}
\hat{g}(i\omega_n;\mathbf{ r},\mathbf{ r}') & \hat{f}(i\omega_n;\mathbf{ r},\mathbf{ r}')\\
\hat{f}^\dag(i\omega_n;\mathbf{ r},\mathbf{ r}') & \nunder[3]{\hat{g}}(i\omega_n;\mathbf{ r},\mathbf{ r}')
\end{array}  \right)
\end{equation}
Here, $4\times 4$ matrices are symbolized by a ``hat'', $\hat{\square}$, and $8\times 8$ matrices by the ``check'' symbol, $\check{\square}$.

\small
\begin {table}[b]
\caption {Band parameters of MBP under the influence of strain\cite{Pereira,Voon1,Voon2}. } \label{table}
\begin{center}
\begin{tabular}{c*{4}{c}c}
\hline
\hline
$u_0$(eV) & $\delta_0$(eV) & $\alpha_x$(eV) & $\alpha_y$(eV) & $\mu_x$(eV)    \\
-0.42  & +0.76  & +3.15  & -0.58  & +2.65     \\
\hline
 $\mu_y$(eV) & $\eta_x$(eV$\cdot \textup{\AA}^2$)  & $\eta_y$(eV$\cdot \textup{\AA}^2$)  & $\gamma_x$(eV$\cdot \textup{\AA}^2$)  & $\gamma_y$(eV$\cdot \textup{\AA}^2$) \\
 +2.16  &  +0.58  & +1.01 & +3.93 & + 3.83  \\
\hline
$\beta_{xx}$(eV$\cdot \textup{\AA}^2$) & $\beta_{yx}$(eV$\cdot \textup{\AA}^2$) & $\beta_{xy}$(eV$\cdot \textup{\AA}^2$) & $\beta_{yy}$(eV$\cdot \textup{\AA}^2$)  \\
-3.48 & -0.57 & +0.80 & +2.39\\
\hline
$\nu_{xx}$(eV$\cdot \textup{\AA}^2$)  & $\nu_{yx}$(eV$\cdot \textup{\AA}^2$) & $\nu_{xy}$(eV$\cdot \textup{\AA}^2$) & $\nu_{yy}$(eV$\cdot \textup{\AA}^2$) & $\chi_y$(eV$\cdot \textup{\AA}$)  \\
-10.90 & -11.33 & -41.40 & -14.80 & +5.25\\
\hline
\hline
\end{tabular}
\end{center}
\end{table}
\normalsize

\begin{figure*}
\includegraphics[width=18.0cm,height=8.30cm]{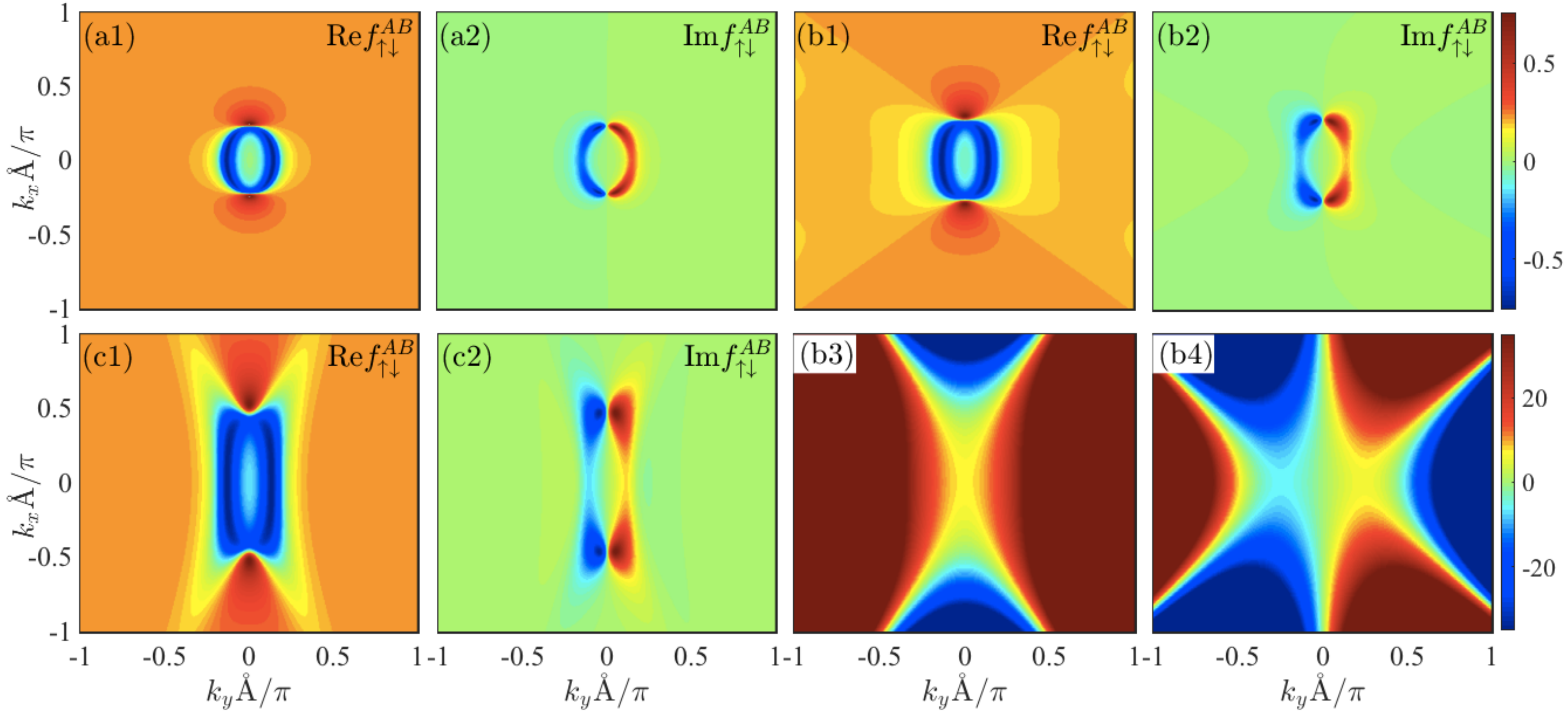}
\caption{\label{fig1} (Color online).
Real and imaginary parts of superconducting pairing correlation $f_{\ua\da}^{AB}$ as a function of momenta $k_x$ and $k_y$ using the material parameters in Table \ref{table}. [(a1) and (a2)] $\varepsilon_{xx}=\varepsilon_{yy}=0$, [(b1) and (b2)] $\varepsilon_{xx}=\varepsilon_{yy}=+0.07$, [(c1) and (c2)], and $\varepsilon_{xx}=\varepsilon_{yy}=+0.15$. Panels (b3) and (b4) are the numerator of $\text{Re}f_{\ua\da}^{AB}$ and $\text{Im}f_{\ua\da}^{AB}$ shown in (b1) and (b2). The chemical potential is fixed at $\mu=2.5$eV and $|\Delta|\neq 0$. The color scale of (a1)-(c2) are given by that of shown next to (b2), while (b3) and (b4) are scaled by the color bar shown next to (b4).}
\end{figure*}

We first set the exchange field to zero, ${\bf h}=0$. Also, in line with experiments, we assume that the system is translationally invariant in the $xy$ plane and thus diagonal in momentum. Consequently, one ends up with a set of algebraic equations in momentum space. After solving for the Green's function, we obtain the following pairing components:
\begin{subequations}\label{gf1}
\begin{eqnarray}
&{\cal D}f_{11}={\cal D}f_{\ua\ua}^{AA}=0,\label{gfa}\\
&{\cal D}f_{12}={\cal D}f_{\ua\da}^{AA}=2 i \Delta  (\chi_y k_y (\mu - \Omega)-\Lambda
   \omega_n),\label{gfb}\\
&{\cal D}f_{13}={\cal D}f_{\ua\ua}^{AB}=0,  \label{gfc}\\
&{\cal D}f_{14}={\cal D}f_{\ua\da}^{AB}=-\Delta  \left(\Delta ^2+(\chi_y k_y-i \Lambda )^2+(\mu
   -\Omega )^2+\omega_n^2\right),~~~~~~\label{gfd}
\end{eqnarray}
\end{subequations}
in which $\cal D$ is given by
\begin{equation}\label{DD}
\begin{split}
{\cal D}=&2 \omega_n^2 (\Lambda ^2+\Delta ^2+\Omega
   ^2)+(\Delta ^2-\Lambda^2+\Omega^2)^2+\\& \chi_y^4 k_y^4+ 2 \chi_y^2 k_y^2 (\Lambda
   ^2+\Delta ^2-\Omega^2+\omega_n^2)+\mu
   ^4+\\& \omega_n^4+ 2 \mu ^2
   (\Delta ^2-\Lambda^2-\chi_y^2
   k_y^2+ 3 \Omega^2+\omega_n^2)-  \\&4
   \mu  \Omega  (\Delta ^2-\Lambda^2-\chi_y^2
   k_y^2+\Omega^2+\omega_n^2)-4 \mu ^3 \Omega,
\end{split}
\end{equation}
and we defined
\begin{subequations}\label{cof}
\begin{gather}
 \Omega=\sum\limits_{i,j=x,y} u_0+\alpha_i\varepsilon_{ii} +(\eta_j+\beta_{ij}\varepsilon_{ii})k_j^2,\label{cofa}\\
 \Lambda=\sum\limits_{i,j=x,y} \delta_0+\mu_i\varepsilon_{ii} +(\gamma_j+\nu_{ij}\varepsilon_{ii})k_j^2.\label{cofb}
\end{gather}
\end{subequations}
\begin{figure*}[t]
\includegraphics[width=18.00cm,height=8.0cm]{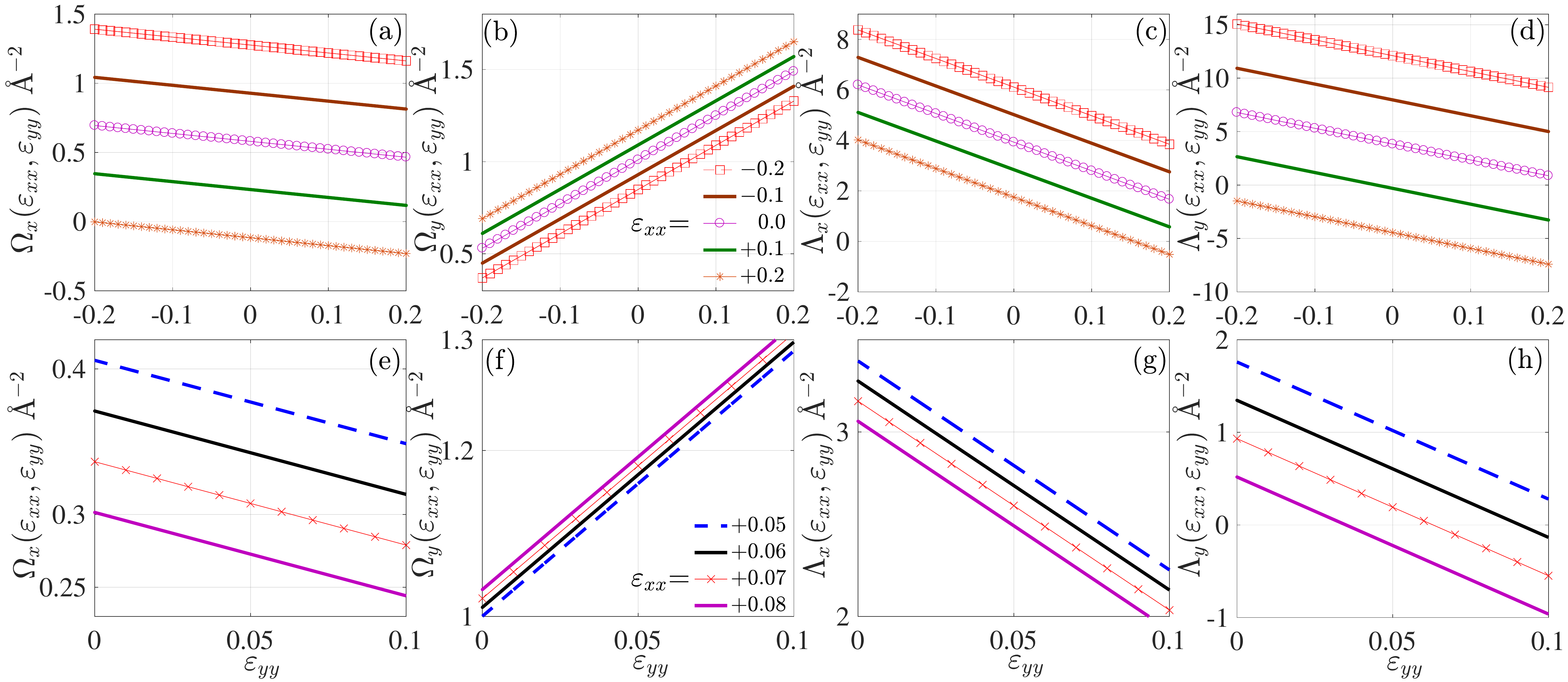}
\caption{\label{fig2} (Color online).
Coefficients of $k_x^2$ and $k_y^2$ in $\Omega$ and $\Lambda$, given in Eq. (\ref{cof}), as a function of strain in the $y$ direction, $\varepsilon_{yy}$, for various values of strain in the $x$ direction, $\varepsilon_{xx}$. Panels (a)/(e) and (b)/(f) show $\Omega_x(\varepsilon_{xx},\varepsilon_{yy})$ and $\Omega_y(\varepsilon_{xx},\varepsilon_{yy})$. Panels (c)/(g) and (d)/(h) exhibit $\Lambda_x(\varepsilon_{xx},\varepsilon_{yy})$ and $\Lambda_y(\varepsilon_{xx},\varepsilon_{yy})$. Top row covers a broad window of strain parameter values while the bottom row illustrates a closer snapshot in strain parameter space where the symmetry transition occurs.}
\end{figure*}
The other components of $\hat{f}$ can be obtained similarly, and can also be inferred from the symmetry relations among the components of the anomalous Green's function. We find that the superconducting correlations with equal spins in both equal-pseudospin and unequal-pseudospin states, i.e., Eqs. (\ref{gfa}) and (\ref{gfc}), vanish. This is consistent with the preliminary assumption we started with, namely, it follows directly from $\Delta^{AB}_{\ua\ua}=\Delta^{AA,BB}_{\ua\ua}=0$. One nonzero superconducting correlation is the equal-pseudospin opposite-spin pairing, i.e., Eq. (\ref{gfb}). This component is purely imaginary, odd in frequency, and odd in parity (i.e., changing $k_x,k_y$ to $-k_x,-k_y$). Thus, $f_{\ua\da}^{AA}$ belongs to the $p$-wave symmetry class. The most important pairing correlation is given in Eq. (\ref{gfd}). This pairing involves unequal-pseudospin opposite-spins and is even in frequency. According to Eqs. (\ref{cof}), both terms $\Omega$ and $\Lambda$ have the form ``$ak_x^2+bk_y^2+c$''. For $\Omega$, we find 
\begin{subequations}\label{cof1}
\begin{eqnarray}
a&=&\eta_x+\beta_{xx}\varepsilon_{xx}+\beta_{yx}\varepsilon_{yy}, \\b&=&\eta_y+\beta_{xy}\varepsilon_{xx}+\beta_{yy}\varepsilon_{yy}, \\c&=&u_0+\alpha_x\varepsilon_{xx}+\alpha_y\varepsilon_{yy}, 
\end{eqnarray}
\end{subequations}
while for $\Lambda$ we have 
\begin{subequations}\label{cof2}
\begin{eqnarray}
a&=&\gamma_x+\nu_{xx}\varepsilon_{xx}+\nu_{yx}\varepsilon_{yy}, \\b&=&\gamma_y+\nu_{xy}\varepsilon_{xx}+\nu_{yy}\varepsilon_{yy}, \\c&=&\delta_0+\mu_x\varepsilon_{xx}+\mu_y\varepsilon_{yy}. 
\end{eqnarray}
\end{subequations}
Our conclusions for the change of pairing correlations' symmetry classes hinge solely on the actual forms of $\Omega$ and $\Lambda$, i.e., ``$ak_x^2+bk_y^2+c$'', and the relative sign of $a/b$. Hence, when $\text{sgn}(a/b)$ is positive, $k_y\Lambda$ and $\Omega$ (and $\Lambda$) belong to $p$- and $s$-wave symmetry classes, respectively. Depending on the applied mechanical strain, $\text{sgn}(a/b)$ can change to negative. Consequently, the leading symmetry classes change to $f$- and $d$-symmetries, as now $\Omega$ and $\Lambda$ are given by 
\begin{equation}
c\pm |a|k_x^2\mp |b|k_y^2,
\end{equation}
(see also the numerical illustrations in the next paragraph). Expanding $(\chi_y k_y-i \Lambda )^2$ in Eq. (\ref{gfd}), we find a term proportional to 
\begin{equation}
\chi_y k_y\Lambda=c\chi_yk_y\pm \chi_yk_y(|a|k_x^2-|b|k_y^2),
\end{equation}
belonging now to $p$- and $f$-wave classes. In a rotated frame by unitary transformation $U=\exp (i\pi\tau_y/4)$, the term $\Lambda$ in the Hamiltonian shall be attached to a $\tau_z$ matrix rather than $\tau_x$. In this case, we find 
\begin{subequations}
\begin{eqnarray}
{\cal D}f_{\ua\da}^{AA}&=&2i\Delta\chi_yk_y(\mu+\Lambda-\Omega),\\
{\cal D}f_{\ua\da}^{AB}&=&-\Delta(\Delta^2+\chi_y^2k_y^2+(\mu-\Omega)^2-(\Lambda+i\omega_n)^2).
\end{eqnarray}
\end{subequations}
Note that in the absence of anisotropic term $\chi_yk_y$, the pairing correlations possess effective $s$-wave symmetry and $f_{\ua\da}^{AA}$ vanishes. Also, the applied strain can modify the band structure as Ref. \onlinecite{alidoust2018bp} and Appendix \ref{apdx} throughly discuss.

The pairing correlation $f_{\ua\da}^{AB}$ is a complicated function of the momenta $k_x$ and $k_y$. Therefore, to determine the effective symmetries of $f_{\ua\da}^{AB}$, we employ realistic values for the band parameters given in Table \ref{table} and numerically evaluate Eq. (\ref{gfd}) at the first Matsubara frequency mode, $n=0$, and a small nonzero temperature. To be specific, we consider three different sets for a biaxial strain (a) $\varepsilon_{xx}=\varepsilon_{yy}=0$ (no strain), (b) $\varepsilon_{xx}=\varepsilon_{yy}=+0.07$, and (c) $\varepsilon_{xx}=\varepsilon_{yy}=+0.15$. The numerical results are shown in Fig. \ref{fig1}. Figures \ref{fig1}(a1)/\ref{fig1}(a2), \ref{fig1}(b1)/\ref{fig1}(b2), and \ref{fig1}(c1)/\ref{fig1}(c2) illustrate the real and imaginary parts of $f_{\ua\da}^{AB}$ for the strain parameter sets (a), (b), and (c), respectively. It is apparent that Fig. \ref{fig1}(a1) has an effective $s$-wave symmetry, which is the real part of $f_{\ua\da}^{AB}$, whereas Fig. \ref{fig1}(a2) shows a $p$-wave symmetry. These observations are consistent with the analyses presented in the previous paragraph. When the strain is increased, an effective symmetry of type $d$-wave is manifested in Figs. \ref{fig1}(b1) and \ref{fig1}(c1). As mentioned above, the superconducting correlation $f_{\ua\da}^{AB}$ is a combination of different symmetries. Nonetheless, as demonstrated by the numerics, an appropriate variation of the biaxial strain can drive the leading superconducting pairing symmetry from one symmetry class to another with prominent signatures. To further shed light on the symmetries, we illustrate the numerator of $\text{Re} f_{\ua\da}^{AB}$ and $\text{Im} f_{\ua\da}^{AB}$, with identical parameter values used in Fig. \ref{fig1}(b), in Figs. \ref{fig1}(b3) and \ref{fig1}(b4). As expected, the real and imaginary parts of numerator exhibit $d$-wave and $f$-wave symmetries, respectively. [The large amplitude of denominator limits the color map's ability to fully illustrate these details in Figs. \ref{fig1}(b1) and \ref{fig1}(b2)]. We have also examined the pairing correlations Eqs. (\ref{gf1}) when the exchange field has components in different directions. Our numerical results (not shown) illustrate that an exchange field in the plane of MBP yields nonzero values for all components of the anomalous Green's function. Conversely, a perpendicular exchange field does not affect the equal spin-triplet components, and only alters Eqs. (\ref{gfb}) and (\ref{gfd}), involving opposite spin-triplet correlations. We find that in all cases, the change of symmetry classes occurs for the spin-triplet correlations as well. 

Having uncovered the symmetry conversion of superconducting pairings, in order to provide a broader view of the parameter space where these symmetry crossovers occur, we have separated out the coefficients of $k_x^2$ and $k_y^2$ in Eq. (\ref{cof}). The coefficients of $k_x^2$ in $\Omega$ and $\Lambda$ are denoted by $\Omega_x(\varepsilon_{xx},\varepsilon_{yy})$ and $\Lambda_x(\varepsilon_{xx},\varepsilon_{yy})$ (identical to coefficients ``$a$'' given above), while the coefficients of $k_y^2$ are shown by $\Omega_y(\varepsilon_{xx},\varepsilon_{yy})$ and $\Lambda_y(\varepsilon_{xx},\varepsilon_{yy})$ (equivalent to coefficients ``$b$'' introduced above). Employing the band parameters given in Table \ref{table}, we have plotted these coefficients by continuously varying $\varepsilon_{yy}$ and few representative values to $\varepsilon_{xx}$ in Fig. \ref{fig2}. For the (a) set of strain, all coefficients are positive, i.e., $\Omega_{x,y}>0, \Lambda_{x,y}>0$ and, hence, we end up with terms of type $|a|k_x^2+|b|k_y^2$ which possess the $s$-wave symmetry class. In contrast, for the set of strain used for obtaining Fig. \ref{fig1}(b), the coefficients have opposite signs, i.e., 
\begin{equation}
\left\{\begin{array}{c}
\Omega_{x}>0, \Omega_{y}>0,\\
\Lambda_{x}>0, \Lambda_{y}<0.
\end{array}\right.
\end{equation}
For the case of higher strain, i.e., $\varepsilon_{xx}=\varepsilon_{yy}=+0.15$ shown in Fig. \ref{fig1}(c), we find 
\begin{equation}
\left\{\begin{array}{c}
\Omega_{x}<0, \Omega_{y}>0,\\
\Lambda_{x}>0, \Lambda_{y}<0.
\end{array}\right.
\end{equation}
These combinations produce terms of types $\pm |a|k_x^2\mp|b|k_y^2$ that have $d$-wave symmetry class. We note that our analyses are independent of the chemical potential as long as MBP develops superconductivity \cite{Shao1}. The external strain can be replaced by other methods, which modify the Green’s function coefficients, for example doping. Depending on the direction of the applied strain, a specific threshold value is needed -as can be extracted from Fig. \ref{fig2}- to achieve the required symmetry change (see also Appendix \ref{apdx}). In general, our findings and predictions are not limited to MBP and can occur in any two- and three-dimensional materials provided that the low-energy Hamiltonian supports similar symmetries as that of MBP, and a sufficiently strong negative ratio between momenta is accessible by some means.

As stated in the Introduction, the optical band gap only is $\sim 1.52$eV that appears in our model Hamiltonian as $\delta_0$ in Table \ref{table}. By accounting for the exciton binding energy, the band gap increases to $\sim 2.3$~eV \cite{V.Tran}. Nevertheless, for our discussions and conclusions, the modification of the band gap from $\sim 1.52$\cite{Rudenko,Pereira} to $\sim 2.3$~eV\cite{V.Tran} has negligible influence because our findings mainly rely on the sign change of the coefficients of momenta $k_x^2$ and $k_y^2$ given in Eqs. (\ref{cof})-(\ref{cof2}).

As the pairing correlations in Eqs. (\ref{gf1}) illustrate, MBP under the influence of strain can potentially host angle dependent pairings ($p$-, $d$-, and $f$-wave), and their combinations. The other main determinative factor for the superconducting gap and its temperature dependency is the interaction potential $V({\textbf{k},\textbf{k}'})$. Accounting for momentum, valley, and spin degrees of freedom, the self-consistent superconducting gap function is given by;
\begin{equation}\label{gapfunc}
\Delta^{\alpha\alpha'}_{\beta\beta'}({\textbf{k}})=-\sum_{\textbf{k}'} V^{\alpha\alpha',\sigma\sigma'}_{\beta\beta',\rho\rho'}({\textbf{k},\textbf{k}'}) f^{\sigma\sigma'}_{\rho\rho'}(\textbf{k}'),
\end{equation}
where $\alpha\alpha', \beta\beta'$ are pseudospin and spin indices. The interaction potential function can be expanded through the spherical harmonics. Thus, for an $s$-wave, $V\propto 1$: for a $d_{x^2-y^2}$-wave, $V\propto k_x^2-k_y^2$: for an $f_{y(3x^2-y^2)}$-wave, $V\propto k_y(3k_x^2-k_y^2)$; and so forth. When an appropriate interaction potential is specified, $\Delta^{\alpha\alpha'}_{\beta\beta'}({\textbf{k}})$ provides an integral equation to find temperature dependence of the superconducting gap. In practice, when the interaction potential $V({\textbf{k},\textbf{k}'})$ supports the symmetries explored through the Green's function, our results suggest that the superconducting critical temperature, initially pertaining to $s$-wave superconductivity, can be significantly altered by applying an appropriate either uniaxial or biaxial strain, as the superconducting pairing symmetry turns into a $d_{ax^2-by^2}$. This prediction can be checked in an experiment. Another experiment that provides distinctive and direct evidence for our prediction of symmetry change is an angular point contact tunneling spectroscopy experiment. Such an experiment determines the angular dependence of superconductivity  before and after applying the biaxial strain, thus differentiating between angle-independent and angle-dependent symmetry classes. An alternative approach that can reveal the symmetry change of superconducting correlations is the Meissner effect. This method may provide a clear and direct image of the symmetry change discussed above \cite{arxiv_nonlinearMSE}. 

In our analyses above, we have simply adopted the BCS scenario of superconductivity in which two particles with opposite spins and momenta are coupled from different pseudospins. We have assumed that the influence of strain in superconducting MBP is encoded in the electron-electron coupling interaction potential so that it changes by the amplitude of this interaction. Nonetheless, if it will be experimentally proven that strain introduces momentum-dependent changes to the electron-electron coupling interaction potential, similar to those discussed above, then one should repeat the above calculations with momentum-dependent potentials. 

First-principles calculations show that BP withstands high strains up to $40\%$ without any rupture or dislocations\cite{Carvalho,wei,Li1}. Nonetheless, we have used low strain values ($7\%,15\%$) and yet find clearly the symmetry conversion. Our inspection of the strained MBP demonstrates that a strain of $>$$6\%$ (biaxial of the type discussed in Fig. \ref{fig1}, i.e., $\varepsilon_{xx}=\varepsilon_{yy}$) initiates the symmetry transformation discussed above. This low strain regime is promising to the feasibility of proposed experiments. At large strains other allotropes may become energetically competitive \cite{struc_change0,struc_change1,struc_change2}. Here we assume that MBP does not undergo a structural transition for the range of strain values used. 

We emphasize that the low strain introduced in our proposal does not close the band gap of MBP and, as discussed extensively above, with reference to Fig. \ref{fig2}, it can be compressive, tensile, or a combination of them both uniaxial or biaxial. In order to provide sufficiently high particle density for superconductivity, we propose the injection of quasiparticles in the system as controlled by the chemical potential $\mu$, here chosen to be $2.5$~eV.

\section{Conclusions}
Exploiting both a Green's function approach in the equilibrium state and the low energy effective Hamiltonian of MBP, we derive the components of anomalous Green's function in the presence of intervalley spin-singlet superconductivity. We show that equal-pseudospin spin-singlet correlations emerge, which are odd in parity. More importantly, making use of realistic values to the band parameters of MBP (obtained through first-principles calculations), our results reveal that MBP under the influence of an appropriate weak strain ($>$$6\%$) can change the symmetry class of $s$-wave and $p$-wave superconducting correlations into $d$-wave and $f$-wave classes, respectively. This finding offers MBP as a fertile platform to devise superconducting pairing symmetry switches with externally adjustable parameters and may advance understanding of the underlying mechanism of some high-$T_c$ superconductors.

\begin{acknowledgments}
M.A. is supported by Iran's National Elites Foundation (INEF). Center for Nanostructured Graphene is supported by the Danish National Research Foundation (Project No. DNRF103).
\end{acknowledgments}

\newpage

\onecolumngrid
\appendix

\section{}\label{apdx}

\renewcommand{\thefigure}{S\arabic{figure}}

\setcounter{figure}{0}

\begin{figure}[t]
\includegraphics[width=12.90cm,height=6.0cm]{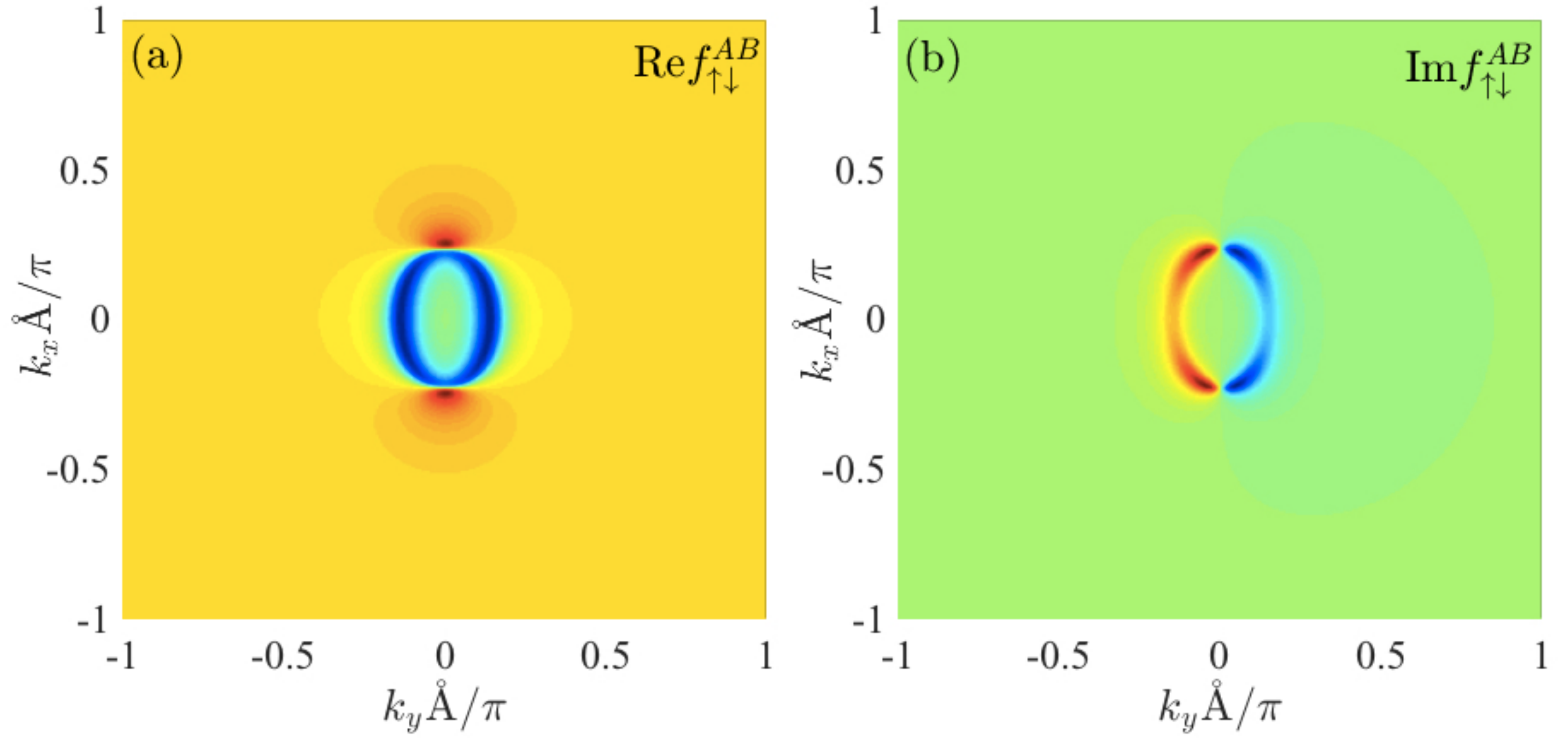}
\caption{\label{s0} (Color online).
Real and imaginary parts of unequal-pseudospin oppostie-spin superconducting pair correlation $f_{\uparrow\downarrow}^{AB}$, where $\varepsilon_{xx}=\varepsilon_{yy} = 2\%$. Compared to Figs. \ref{fig1}(a1) and \ref{fig1}(a2) of the main text where $\varepsilon_{xx}=\varepsilon_{yy} = 0\%$, we see no significant change in the symmetry of $f_{\uparrow\downarrow}^{AB}$. As it is analyzed in the main text and its Fig. \ref{fig2} illustrates, the symmetry change occurs at strains larger than $6\%$.
}
\end{figure}

In this Appendix we present more results for the intervalley unequal-spin superconducting correlation under strain. We also study the influence of $\varepsilon_{xx}$, $\varepsilon_{yy}$ on the low energy band structure both in confined and infinite systems.

In order to demonstrate that the strain-induced symmetry change explored in the main text is not accidental, we plot the unequal-pseudospin opposite-spin superconducting correlation $f_{\uparrow\downarrow}^{AB}$ for a representative low value of biaxial strain, i.e., $\varepsilon_{xx}=\varepsilon_{yy} = 2\%$ in Fig. \ref{s0}. Parameter values are otherwise the same as those used to obtain Fig. \ref{fig1}. Compared to the case with zero strain discussed in main text, we observe no significant change in the symmetry of $f_{\uparrow\downarrow}^{AB}$. As stated in the main text, to alter effectively the symmetry of $f_{\uparrow\downarrow}^{AB}$, one needs to exert strains larger than a threshold value regardless of being either uniaxial or biaxial. These threshold values can be extracted from Fig. \ref{fig2}. Also, it is worth mentioning that the discussed strain as a tool to drive the symmetry transformations in this paper might be replaced by other means. One example can be the intercalation of different elements into the parent atoms of monolayer black phosphorus. According to our analyses presented in this paper, one only needs to drive a  sufficiently strong negative ratio between the coefficients of momenta in monolayer black phosphorus.  

Although the band parameters obtained through DFT calculations may change slightly, depending on the underlying first-principles method used, we would like to emphasize that the presented analyses in our work are independent of the band parameters. We clarify the generic conditions under which the transformations discussed in the symmetry profile of superconducting pair correlations are guaranteed. We also refer to a recent work where a good agreement between the first- principles calculations and experimental observations in black phosphorus subject to strain was reported \cite{exp_theor}.

\newpage

\begin{figure*}[t]
\includegraphics[width=18.0cm,height=4.40cm]{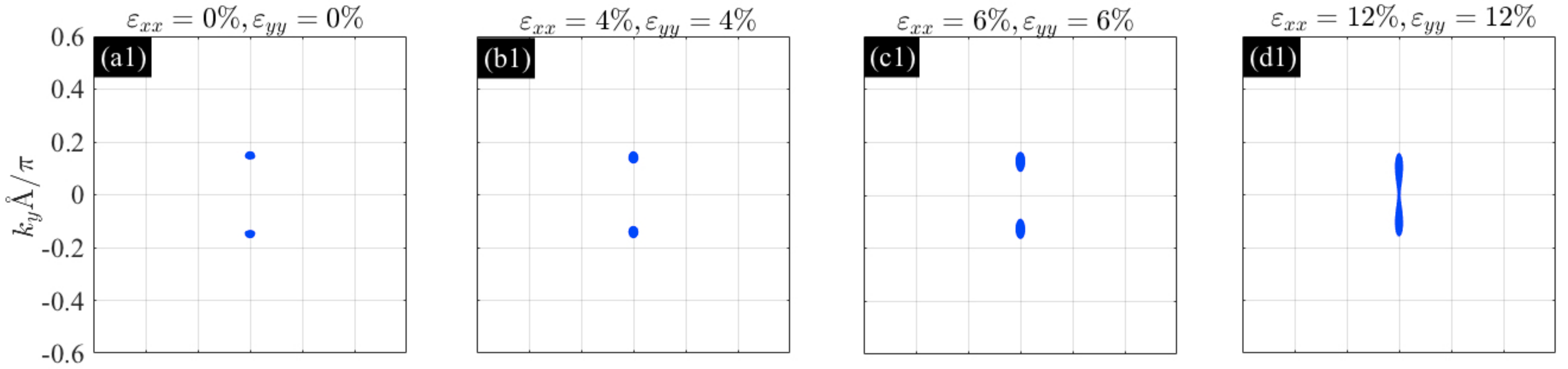}
\includegraphics[width=18.0cm,height=4.0cm]{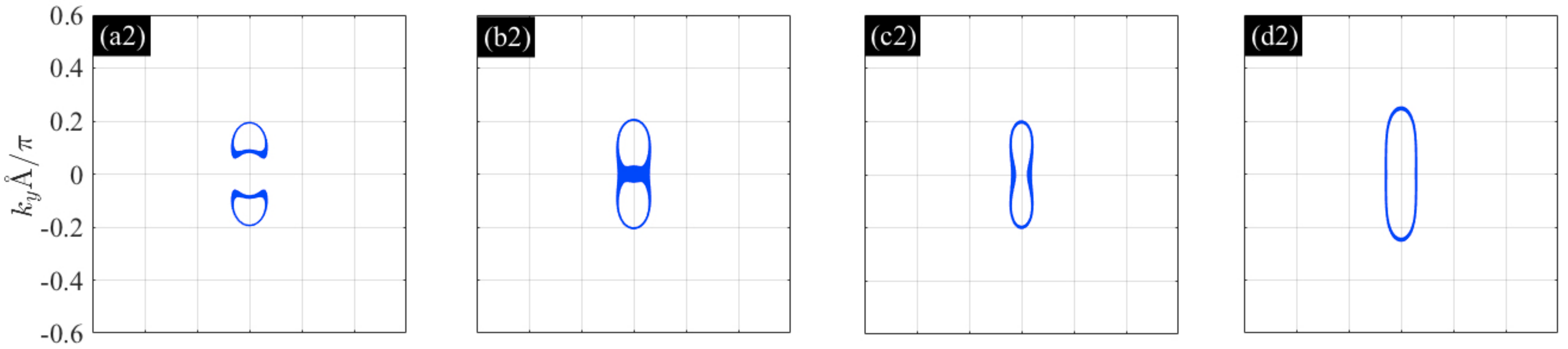}
\includegraphics[ width=18.0cm,height=4.0cm]{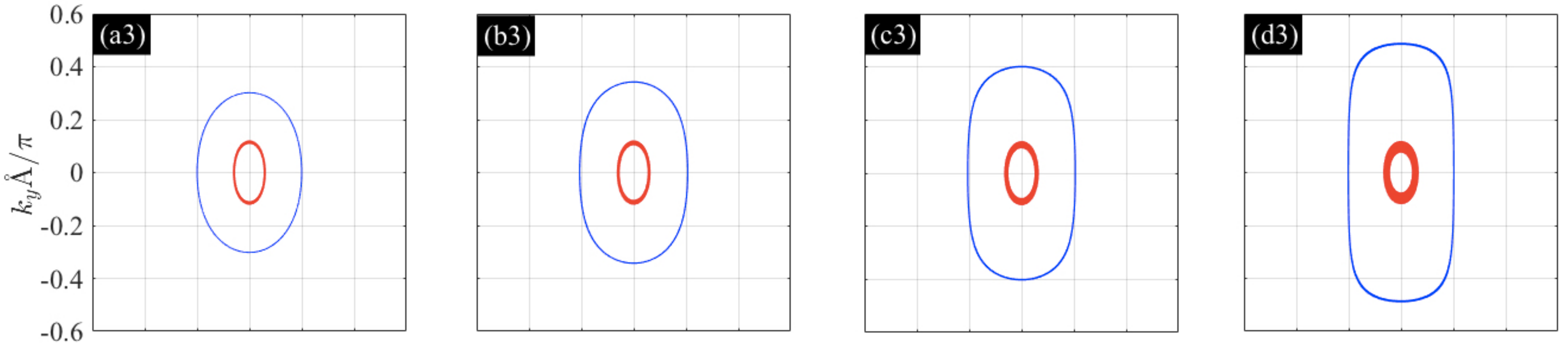}
\includegraphics[ width=18.120cm,height=4.40cm]{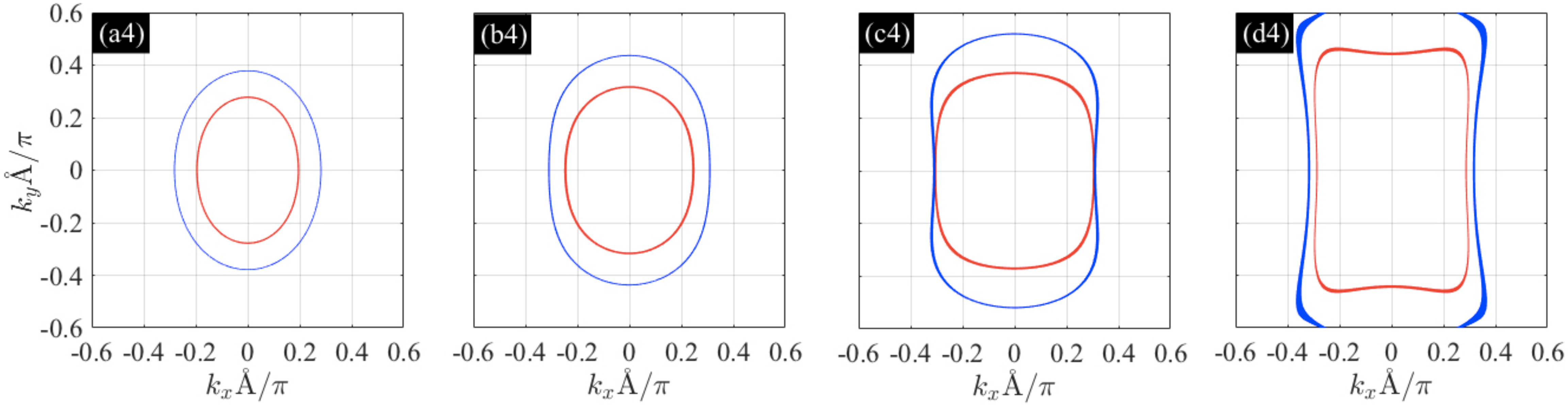}
\caption{\label{s1} (Color online). Isoenergy band structure of monolayer black phosphorus with infinite sizes. In each column we apply a set of representative biaxial strains: (a) $\varepsilon_{xx}=\varepsilon_{yy}=0\%$, (b) $\varepsilon_{xx}=\varepsilon_{yy}=4\%$, (c) $\varepsilon_{xx}=\varepsilon_{yy}=6\%$, and (d) $\varepsilon_{xx}=\varepsilon_{yy}=12\%$. In each row, we set different values for the energy: first row $E=0.15$~eV, second row $E=0.7$~eV, third row $E=3.0$~eV, and fourth row $E=5.0$~eV. }
\end{figure*}
\begin{figure*}[t]
\includegraphics[width=18.0cm,height=5.80cm]{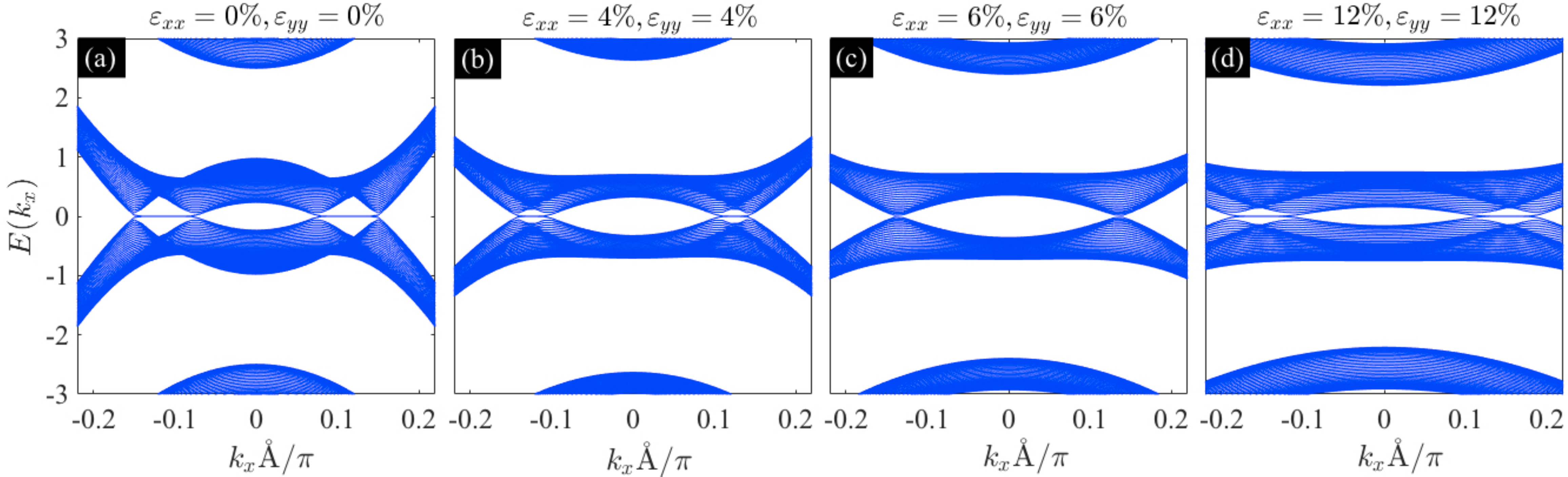}
\caption{\label{s2} (Color online). Band structures of a superconducting monolayer black phosphorus subject to different amounts of uniaxial strain. The system is finite sized in the $y$ direction in line with Ref. \onlinecite{alidoust2018bp}. From left to right, we increase the uniaxial strain: (a) $\varepsilon_{xx}=\varepsilon_{yy}=0\%$, (b) $\varepsilon_{xx}=\varepsilon_{yy}=4\%$, (c) $\varepsilon_{xx}=\varepsilon_{yy}=6\%$, and (d) $\varepsilon_{xx}=\varepsilon_{yy}=12\%$. }
\end{figure*}

It is obvious that the application of strain modifies the band structure of system. To gain insight on how strain influences the band structure, we have considered an infinite sample of superconducting monolayer black phosphorus where momenta are good quantum numbers. This situation allows us to obtain isoenergy contour plots as shown in Figure \ref{s1}. To present a comprehensive view, we have considered four different values of biaxial strain: (a) $\varepsilon_{xx}=\varepsilon_{yy}=0\%$, (b) $\varepsilon_{xx}=\varepsilon_{yy}=4\%$, (c) $\varepsilon_{xx}=\varepsilon_{yy}=6\%$, and (d) $\varepsilon_{xx}=\varepsilon_{yy}=12\%$. In each row, we set a fix value to energy, from top to bottom: $E=0.15, 0.7, 3.0, 5.0$~eV. The other parameters are identical to those used to obtain Fig.\ref{fig1}. At low energies, the band structure contains two Dirac points shown by two solid circles. By increasing the applied biaxial strain, the slope of the bands forming the Dirac points increases and more states move below $E=0.15$~eV so that when $\varepsilon_{xx}=\varepsilon_{yy}=12\%$ the edges of the two Dirac cones touch each other. The band structure shows similar behavior at $E=0.7$~eV up to $\varepsilon_{xx}=\varepsilon_{yy}=4\%$. For larger strain values, the touching point moves away from $E=0.7$~eV, and therefore we see only a nonideal ellipsoid and with further increase of the strain it becomes more rounded. At higher energy values, i.e., $E=3.0$~eV the second band emerges (shown by red). We see that even at zero strain the band structure is not symmetric with respect to momenta rather it generates an ellipsoid in the isoenergy contour plot. By increasing the applied strain, this nonsymmetric behavior is more pronounced so that at $\varepsilon_{xx}=\varepsilon_{yy}=12\%$ the outer band reshapes to a nonideal square. This is highly pronounced at higher energies as shown in the last row of Fig. \ref{s1}. It is clearly seen for the higher strain case that the band structure is warped and exhibits nontrivial modifications to the previous cases with lower strain values.

When the superconducting monolayer black phosphorus is confined, the band structure gains nontrivial features at low energies. To illustrate how the low energy band structure is modified in a finite size system subject to strain, we consider a situation where the system is confined in the $y$ direction with length $d_S=50$~nm. In this case, momentum in the $x$ direction is invariant and allows us to plot the band structure vs $k_x$. Figure \ref{s2} illustrates the influence of strain on Weyl nodes at low energies. In the absence of strain four Weyl nodes exist symmetrically distributed around $k_x=0$. The Weyl nodes in negative and positive $k_x$ are connected by dispersionless flat bands (see Ref. \onlinecite{alidoust2018bp}). The wavefunctions associated with these flat bands are localized at the edges of sample that are known as ``Majorana zero energy modes''. When the strain is increased, the Weyl nodes come closer together, at strains around $\varepsilon_{xx}=\varepsilon_{yy}=7\%$ merge, and, finally, for larger values of strain exchange their locations as seen in Fig. \ref{s2}(d). For further investigations, we refer to Ref. \onlinecite{alidoust2018bp}.

\twocolumngrid

\end{document}